\documentclass[aps,prd,floatfix,showkeys,nofootinbib]{revtex4}
\usepackage[colorlinks=false, urlcolor=blue, pdfborder={0 0 0}]{hyperref}
\usepackage{hyperref}
\pdfoutput=1
\usepackage{amssymb} \usepackage{amsmath} \usepackage{graphicx}
\usepackage{epsfig,latexsym}
\usepackage{ulem}
\usepackage{color}

\newcommand{\AddrAHEP}{
  AHEP Group, Instituto de F\'{\i}sica Corpuscular -- C.S.I.C./Universitat de Val{\`e}ncia \\
  c/ Catedr\'atico Jos\'e Beltr\'an, 2, E- 46980 Paterna
  (Val{\`e}ncia) Spain }
\newcommand{\AddrAQUILA}{
Dipartimento di Fisica,
 Universit\`a di L'Aquila, 67010 Coppito AQ and
LNGS, Laboratori Nazionali del Gran Sasso, 67010 Assergi AQ, Italy}

\def\bef{\begin{figure}}
\def\eef{\end{figure}}

\newcommand{\be}[1]{\begin{equation}\label{#1}}
\newcommand{\beq}{\begin{equation}}
\newcommand{\ee}{\end{equation}}
\newcommand{\beqn}[1]{\begin{eqnarray}\label{#1}}
\newcommand{\eeqn}{\end{eqnarray}}
\newcommand{\bd}{\begin{displaymath}}
\newcommand{\ed}{\end{displaymath}}

\def\lsim{\raise0.3ex\hbox{$\;<$\kern-0.75em\raise-1.1ex
e\hbox{$\sim\;$}}}
\def\gsim{\raise0.3ex\hbox{$\;>$\kern-0.75em\raise-1.1ex
\hbox{$\sim\;$}}}
\def\simlt{\mathrel{\lower2.5pt\vbox{\lineskip=0pt\baselineskip=0pt
           \hbox{$<$}\hbox{$\sim$}}}}
\def\simgt{\mathrel{\lower2.5pt\vbox{\lineskip=0pt\baselineskip=0pt
           \hbox{$>$}\hbox{$\sim$}}}}
\def\unity{{\hbox{1\kern-.8mm l}}}

\renewcommand{\to}{\rightarrow}

\renewcommand{\to}{\rightarrow}

\def\lsim{\mathrel{\mathop  {\hbox{\lower0.5ex\hbox{$\sim$}
\kern-0.8em\lower-0.7ex\hbox{$<$}}}}}
\def\gsim{\mathrel{\mathop  {\hbox{\lower0.5ex\hbox{$\sim$}
\kern-0.8em\lower-0.7ex\hbox{$>$}}}}}

\newcommand{\SM}{{standard model }}
\def\sm{ $\mathrm{SU(3)_c \otimes SU(2)_L \otimes U(1)_Y}$}
\def\u3u311{$\mathrm{U(3)_c \otimes U(3)_L \otimes U(1) \otimes U(1)' }$ }
\def\s3s31{$\mathrm{SU(3)_c \otimes SU(3)_L \otimes U(1)_X}$ }
\def\lfv{lepton flavour violation }

\begin{document}
\title{String completion of an \s3s31 electroweak model}

\author{Andrea Addazi} \email{andrea.addazi@infn.lngs.it}
\affiliation{\AddrAQUILA} 

\author{J.~W.~F.~Valle} \email{valle@ific.uv.es}
\author{C.A. Vaquera-Araujo} \email{vaquera@ific.uv.es}
\affiliation{\AddrAHEP} \pacs{} 


\begin{abstract}

  The extended electroweak \s3s31 symmetry framework ``explaining''
  the number of fermion families is revisited.
  While $331$-based schemes can not easily be unified within the
  conventional field theory sense, we show how to do it within an
  approach based on D-branes and (un)oriented open strings, on
  Calabi-Yau singularities.
  We show how the theory can be UV--completed in a quiver setup, free
  of gauge and string anomalies.  Lepton and baryon numbers are
  perturbatively conserved, so neutrinos are Dirac--type, and their
  lightness results from a novel TeV scale seesaw mechanism. 
  Dynamical violation of baryon number by exotic instantons
  could induce neutron-antineutron oscillations, with proton decay
  and other dangerous R--parity violating processes
  strictly forbidden.

\end{abstract}

\maketitle

\section{Introduction}

Among the open challenges in the \SM we encounter issues like:
Why we have three species of fermions? 
Why the neutrino masses are so small?
Why fundamental couplings unify?
How is gravity incorporated in a fundamental way?
One of the early extended electroweak models based on the \s3s31 gauge
group ~\cite{Singer:1980sw,valle:1983dk} ``explains'' the number of
fermion families from the requirement of anomaly cancellation.
Indeed the theory is anomaly free if and only if the number of quark
colors is equal to the number of families, i.e. three (species of
fermions) is related to quantum consistency~\cite{Singer:1980sw,valle:1983dk,Foot:1994ym,Frampton:1992wt,Pisano:1991ee,Hoang:1995vq,Tully:2000kk}.
Recently this scenario has been revamped in order to also provide a
framework for naturally light neutrinos without invoking superheavy
physics~\cite{Boucenna:2014ela}. In this scheme these two fundamental
issues get related through the embedding of the \SM gauge group in
\s3s31. 
In the simplest 3-3-1 model considered recently neutrino masses were
radiatively generated by one-loop corrections, involving new neutral
gauge bosons associated to lepton number violating interactions
\cite{Boucenna:2014ela}.
Within a simple variant it has been shown that the same physics
involved in small neutrino mass generation may also achieve gauge
coupling unification~\cite{Boucenna:2014dia}, alternative to
conventional grand unified theories.

A drawback of such \s3s31 based-models is, however, that they cannot
be easily embedded in a conventional Grand Unified Theory (GUT). In
order to achieve gauge coupling unification the authors in
Ref.~\cite{Boucenna:2014dia} considered an alternative more
complicated version of the model, in which the presence of a neutral
sequential lepton octet allowed for the merging of the gauge couplings
at high energies in the absence of a \textit{bona fide} grand-unified
structure. 
A more ambitious theoretical question is whether such a structure
could be UV--completed and understood in more fundamental terms.

Here we show that our desire of obtaining a consistent string
completion of this type of 3-3-1 theories leads us to the novel
variety of seesaw mechanism proposed in~\cite{Valle:2016kyz}, in which
neutrinos are Dirac particles with masses generated at the tree
level.
Moreover, we find that neutrino masses vanish in the limit where the
up-quark mass vanish. As a result, consistency of the neutrino sector
with the observed quark masses suggests that the new dynamics
associated with neutrino mass generation must reside near the TeV
scale.
On this basis we expect that this model can be directly tested at LHC
in the next run, through the resonant production of new fields
involved in the neutrino mass generation mechanism.
In particular, a new $Z'$ boson can be produced through the Drell-Yan
processes.  This boson couples to \SM particles and to the new
isosinglet neutral leptons \cite{Deppisch:2013cya}.
Another interesting indirect signature predicted by this model is
$b\to s\mu^{+}\mu^{-}$, gauge mediated by the new $Z'$
boson~\cite{Buras:2014yna}. Likewise, there are also \lfv signals,
recently investigated in a simple variant of these
models~\cite{Boucenna:2015zwa}.

\section{String considerations}
\label{sec:string-cons}

In this paper, we show how a \s3s31 model can be embedded into an open
string theory. In particular, we will show how this model can be
obtained by a system of intersecting D-branes and open (un)oriented
strings attached to D-branes.
The open string models~\footnote{For general aspects open-strings
  theories see
  \cite{Sagnotti:1987tw,Pradisi:1988xd,Bianchi:1990yu,Bianchi:1990tb,Bianchi:1991eu,Angelantonj:1996uy,Angelantonj:1996mw}.}
that can realistically embed \SM-like theories or their extensions can
be divided into three classes:
i) (un)oriented type IIA, with intersecting D6-branes wrapping
3-cycles on the Calabi-Yau compactification $CY_{3}$;
ii) (un)oriented type IIB, with D7-branes and D3-branes wrapping
holomorphic divisors in $CY_{3}$.
iii) type I, with magnetized D9-branes wrapping a $CY_{3}$.
Here we will focus on the first class.  In this case, we can directly
calculate low energy interactions for $\alpha_{s}\to 0$,
obtaining just an ${\cal N}=1$ supergravity coupled with matter
fields.
In particular, we will discuss a simple example of a ``quiver field
theory'' embedding of \s3s31 locally free of stringy
anomalies or tadpoles.

In general, a ``quiver'' is simply a diagrammatic representation of a
gauge theory.  A supersymmetric quiver (as in our case) includes all
the matter (super)field content, represented with arrows, and their
interactions.  The corresponding diagrams have the following
conventions \footnote{See \cite{Bianchi:2013gka} for a useful papers
  on (un)oriented quivers' technology.}:
i) gauge groups are nodes, which are in correspondence with the gauge
superfields;
ii)  superfields are oriented lines between nodes;
iii) superfields in the adjoint representations are arrows going in
and out on the same node; those in the bi-fundamental representations
$({\bf M}, {\bf \bar{P}})$ or $({\bf \bar{M}}, {\bf {P}})$ link two
different nodes/gauge groups;
iv) the number of arrows on a line corresponds to the multiplicity of
the same superfield;
v) Closed oriented paths (arrows with the same orientation) like
triangles, quadrangles, and so on, represent possible gauge-invariant
interaction terms in the superpotential.

 In open string theories, quiver diagrams are particularly powerful.
 This is because D-brane dynamics on Calabi-Yau singularities is
 described by quiver field theories in the low energy limit
 $\alpha_{s}\to 0$.  In string theory, lines connecting nodes
 correspond to (un)oriented open strings, while nodes are D-brane
 stacks.
 Intriguingly, we will show how the quiver field theory suggests the
 existence of novel phenomena characteristically ``stringy'' in
 nature.  In particular, we will see how the presence of new anomalous
 massive bosons is inevitably predicted.  In gauge theories, anomalous
 $\mathrm{U(1)}$s lead to quantum inconsistencies, but in string
 theories these can be cured through a Generalized Green-Schwarz
 mechanism (GGS) and Generalized Chern Simon terms.  As a result,
 anomaly cancellation implies mixing vertices connecting the
 $\gamma\,,Z\,,Z'\,,X$ gauge bosons with those of the anomalous
 $\mathrm{U(1)}$s
 \cite{Coriano':2005js,Anastasopoulos:2006cz,Anastasopoulos:2008jt}.
 
 There are other interesting features of quiver field theories related to
 non-perturbative stringy effects which could manifest at low
 energies.  For example, at the low energy limit, a quiver field
 theory admits the
 presence of extra non-perturbative couplings in the effective action,
 generated by "exotic stringy instantons". 
 
 All gauge instantons are classified by the
 Atiyah, Drinfeld, Hitchin, Manin (ADHM) construction. In (un-)oriented type IIA,
 gauge instantons can be described by Euclidean D2 branes (or E2 branes)
 wrapping the {\it same} 3-cycles of ``ordinary physical" D6-branes on
 the Calabi-Yau $CY_{3}$ \cite{Bianchi:2009ij,Bianchi:2012ud}.   
 In (un-)oriented IIB, E-instantons are E3-branes or
 E(-1)-branes wrapping the same holomorphic divisor of ``ordinary
 physical" D7-branes. Furthermore, in type I, E-instantons are
 E5 branes living in the internal space and having the same
 magnetization of the physical D9-branes. (D9-branes wrap on all the
 $CY_{3}$).
 However, in string theory there is another large class of new
 instantons defined as ``exotic''.  They do not exist in gauge
 theories, and do not need to satisfy ADHM constraints.  In type IIA,
 exotic instantons are simply E2-branes wrapping {\it different}
 3-cycles from ordinary physical D6-branes.
 More precisely, exotic instantons have 8 mixed Neumann-Dirichlet
 directions, in contrast to the 4 mixed directions of the D6/D2 case,
 which admits an AHDM construction.  In general, stringy instantons
 lack bosonic zero-modes in the mixed sector. Thus, upon integration
 over the charged Grassmannian moduli, their contribution to the
 superpotential gives rise to positive powers of the fields, opposite
 with respect to the behavior of standard gauge instantons. 

 In quiver field theories, E-brane instantons are represented as
 triangles.  Open strings attached to one ordinary D-brane and one
 E-brane are fermionic moduli or modulini, corresponding to `dotted'
 arrows.  Closed triangles of lines and dotted lines correspond to
 effective interactions among ordinary fields and modulini.
 Integrating out moduli, new effective interactions among ordinary
 fields are generated. As shown in
 \cite{Ibanez:2006da,Ibanez:2007rs,Blumenhagen:2006xt,Cvetic:2007zza,Cvetic:2007ku,Addazi:2014ila,Addazi:2015ata,Addazi:2015rwa,Addazi:2015hka,Addazi:2015fua,Addazi:2015oba,Addazi:2015goa,Addazi:2015yna},
 these new interactions can dynamically violate Baryon and Lepton
 numbers.  Indeed, we will discuss how exotic instantons can directly
 generate $\Delta B=2$ six quarks transitions generating
 neutron-antineutron oscillations.  However, even if $B$ number is violated in our model, the selection rule $\Delta B=2$ emerges
 dynamically, so that proton stability and R parity conservation are
 ensured. 

\section{Un-oriented quiver theory for a \s3s31 Model}

\subsection{Setup}

\begin{figure}[t]
\centerline{ \includegraphics [width=0.9 \columnwidth]{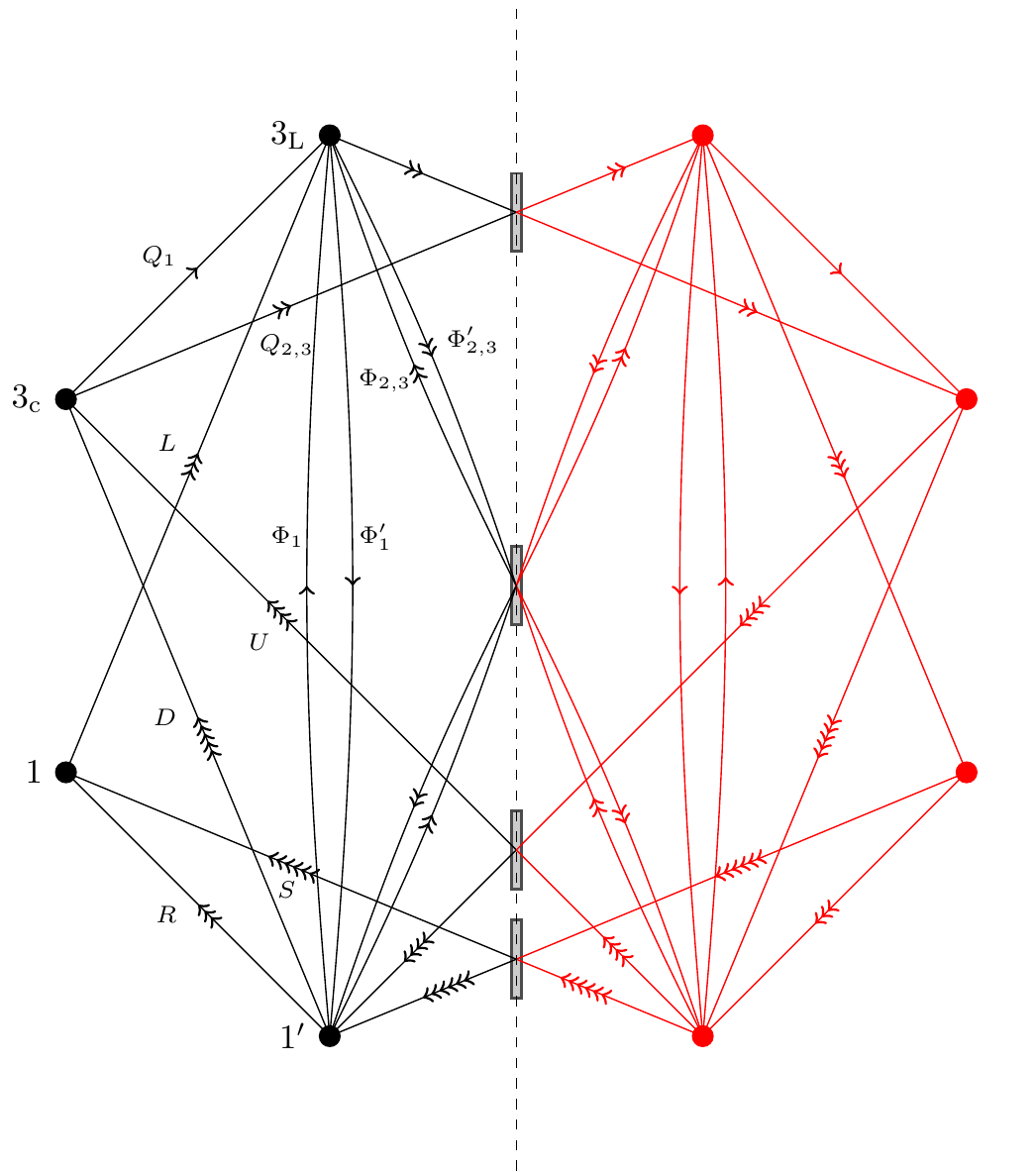}}
\vspace*{-1ex}
\caption{ \u3u311 (un)oriented quiver theory, in the presence of a
  $\Omega$-plane. }
\label{plot}   
\end{figure}

In this section we describe the basic ingredients of our model,
diagrammatically represented as the quiver in Fig. \ref{plot}. 
 This
quiver generates a $\mathcal{N}=1$ supersymmetric theory with gauge
group \u3u311, containing a \s3s31 $\equiv \rm G_{331}$ model.  The
matter content and the (super)field transformation properties under
$(\mathrm{SU(3)_c \,, SU(3)_L)_{X}}$ are
\be{content}
\begin{gathered}
L_{1,2,3}=(1,\bar{3})_{-1/3}\,,\qquad Q_{1}=(3,\bar{3})_{1/3}\,,\qquad Q_{2,3}=(3,3)_{0}\,,\\
R_{1,2,3}=(1,1)_{1}\,,\qquad U_{1\cdots 4}=(\bar{3},1)_{-2/3}\,,\qquad D_{1\cdots 5}=(\bar{3},1)_{1/3},\,\qquad S_{1\cdots 6}=(1,1)_{0}\,,\\
\Phi_{1}=(1,\bar{3})_{2/3}\,,\qquad\Phi_{2,3}=(1,\bar{3})_{-1/3}\,,\qquad\Phi'_{1}=(1,3)_{-2/3},\,\,\,\Phi'_{2,3}=(1,3)_{1/3}\,.
\end{gathered}
\ee
Here $L_{i}$ ($i=1,2,3$) accommodates the $\rm SU(2)_L$ lepton
doublets $\ell_{i}=(E_{L},\nu_{L})^{T}_{i}$ together with new neutral
components $N_{L}{}_{i}$ into the anti-triplet representation of $\rm
U(3)_L$ ; $Q_{i}$ includes the LH doublets
$q_{i}=(u_{L},d_{L})^{T}_{i}$ plus extra quark fields
$u'_{L},s'_{L},b'_{L}$; $R$ stands for the right-handed charged lepton
multiplets and $U$, $D$ contain the right-handed quarks $u_{R},d_{R}$
plus extra three (super)quarks $u'_{R},s'_{R},b'_{R}$. In addition,
there are six $\rm G_{331}$ singlets denoted by $S$, i.e. there is a
pair of gauge singlets in each generation. Finally, the scalar
components of the Higgs superfields $\Phi_{1,2,3},\Phi'_{1,2,3}$ are
responsible for the $\rm G_{331}$ spontaneous symmetry breaking.

The effective trilinear quark and lepton superpotentials,
perturbatively generated, are given as
 \be{super}
 \begin{gathered}
\mathcal{W}_{\rm leptons}=y_{1}LR\Phi_{1}'+y_{2}LS\Phi_{2}'+y_{3}LS\Phi_{3}'\,,\\
\mathcal{W}_{\rm quarks}=y_{4}Q_{1}D\Phi_{1}'+y_{5}Q_{1}U\Phi_{2}'+y_{6}Q_{1}U\Phi_{3}'+y_{7}Q_{2,3}U\Phi_{1}+y_{8}Q_{2,3}D\Phi_{2}+y_{9}Q_{2,3}D\Phi_{3}\,.
\end{gathered}
 \ee
 Each term corresponds to a closed oriented triangle following the
 arrows associated to chiral superfields depicted in Fig. \ref{plot}.
 Moreover, one can see that R-parity violating terms like $LQD$ are
 automatically forbidden at the perturbative level.  This is related
 to the quiver orientations: {\it there are no closed oriented
   triangles corresponding to R-parity violating superpotential
   terms}.
 As a consequence, R-parity is not imposed \textit{ad hoc} in our
 model, but appears as an {\it accidental} symmetry.

 The first stage of gauge symmetry breaking pattern involves a
 Stueckelberg mechanism~\cite{Stueckelberg:1900zz}, while the latter
 is induced by scalar vacuum expectation values (VEVs) through the
 Higgs mechanism. The pattern is
\u3u311 $\to$ \s3s31  $\to$ $\rm G_{\rm SM}$.
Note that the quiver also generates perturbatively the $\mu$-terms for
the Higgs superfields, required for electroweak breaking
\be{mu}
\mathcal{W}_{\mu}=\mu_{1}\Phi_{1} \Phi'_{1}+\mu_{2}\Phi_{2} \Phi'_{2}+\mu_{3}\Phi_{3} \Phi'_{3}+\mu_{4}\Phi_{2} \Phi'_{3}+\mu_{5}\Phi_{3} \Phi'_{2}.
\ee
These terms correspond to closed circuits involving Higgs superfields
in Fig. \ref{plot}.
The proposed quiver can be interpreted as the UV completion of a
particular $\rm G_{331}$ model with extra neutral leptons in the
triplet representation. 
For recent studies in this class of models see for example
Ref.~\cite{Boucenna:2014ela} and
\cite{Boucenna:2015zwa,Valle:2016kyz}. 
A remarkable feature of this quiver construction is that neutrinos are
of Dirac nature, thanks to the presence of a (sequential) pair of
lepton singlets $S$~\footnote{The first feature resembles the
  assumption made in Ref.~\cite{Mohapatra:1986ks}.}
and to the symmetry structure of the model.

\subsection{ Tadpole cancellation and $\mathrm{U(1)_{X}}$ conditions}

The quiver in Fig.\ref{plot} preserves a linear combination
$\mathrm{U(1)_{X}}=\sum_{a}C_{a}\mathrm{U}(1)_{a}$, with
$a=3_{\mathrm{c}},3_{\mathrm{L}},1,1'$ and $\mathrm{
  U(1)_{3_{c}}\subset U(3)_{c}\,,\, U(1)_{3_{L}}\subset U(3)_{L}\,,\,
  U(1)_1\equiv U(1)\,,\, U(1)_{1'}\equiv U(1)'\,,}$ that can be
obtained from the following system: \be{Xsyst}
 \begin{gathered}
X(Q_{1})=C_{\rm 3_{c}}-C_{\rm 3_{L}}=1/3\,,\qquad X(Q_{2,3})=C_{\rm 3_{c}}+C_{\rm 3_{L}}=0\,,\qquad  X(L)=-C_{\rm 3_{L}}+C_{\rm 1}=-1/3\,,\\
X(U)=-C_{\rm 3_{c}}-C_{\rm 1'}=-2/3\,,\qquad X(D)=-C_{\rm 3_{c}}+C_{\rm 1'}=1/3\,,\\
X(R)=-C_{\rm 1}+C_{\rm 1'}=1\,,\qquad X(S)=-C_{\rm 1}-C_{\rm 1'}=0\,,\\
X(\Phi_{1})=-X(\Phi'_{1})=-C_{\rm 3_{L}}+C_{\rm 1'}=\frac{2}{3}\,,\\
X(\Phi_{2,3})=-X(\Phi'_{2,3})=-C_{\rm 3_{L}}-C_{\rm 1'}=-\frac{1}{3}\,.
\end{gathered}
\ee Here we have adopted the convention $+$ for outgoing arrows and
$-$ for incoming ones.
The solution
\be{comb}
\mathrm{ U(1)_{X}=\frac{1}{6}U(1)_{3_{c}}-\frac{1}{6}U(1)_{3_{L}}-\frac{1}{2}U(1)+\frac{1}{2}U'(1)}.
\ee 
corresponds to the defining symmetry of the \s3s31 model~\cite{Valle:2016kyz}.

In order to describe a consistent model, the quiver must be free of
chiral gauge and gravitational anomalies, with $\mathrm{ U(1)_{X}}$
unbroken at the string level.  These requirements are related to the
fulfillment of two more stringent conditions. The first one is local
tadpole cancellation
\cite{Aldazabal:2000dg,Cvetic:2009yh,Cvetic:2011iq}:
\be{condition1}
\sum_{a}N_a(\pi_{a}+\pi_{a'})=4\pi_{\Omega}\,,
\ee
where $\pi_{a}$ denotes 3-cycles wrapped by ``ordinary'' D6-branes, $\pi_{\hat{a}}$ stands for the corresponding 3-cycles wrapped by the
``$\Omega$-image'' D6-branes, and $\pi_{\Omega}$ is the contribution
of the $\Omega$-plane.  More conveniently, Eq.~(\ref{condition1}) can be
expressed in terms of superfields as
\be{condition1a}
\forall \: a\neq a' \qquad \#F_a-\#\bar{F}_a+(N_{a}+4)(\# S_{a}-\# \bar{S}_{a})+(N_{a}-4)(\# A_{a}-\# \bar{A}_{a})=0\,,\ee
with $F_a,\bar{F}_a,S_{a},\bar{S}_{a}, A_{a}, \bar{A}_{a}$ as
fundamental, symmetric and antisymmetric representations of
$\mathrm{U}(N_{a})$ (together with their conjugates).  
Eqs.(\ref{condition1},\ref{condition1a}) are only locally equivalent, and they can not strictly be identified in a global analysis. In the following, we study only the local consistency conditions of our model.
Notice that for
$N_{a}>1$ the above relations coincide with the absence of irreducible
$\mathrm{SU}(N_{a})^{3}$ triangle anomalies~\cite{Singer:1980sw}. The
most important cases are those satisfying $N_a=1$, and can be
interpreted as stringy conditions related to the absence of
`irreducible' $\mathrm{U}(1)^{3}$, {\it i.e.} contributions that arise
from diagrams with insertions of the same $\mathrm{U}(1)$ vector boson
on the same boundary.  The explicit tadpole cancellation follows from
\be{tpc}
\begin{split}
\mathrm{U(3)_c} &:\qquad 3n_{Q_1}+3n_{Q_{2,3}}-n_{D}-n_{U}=3+6-5-4=0\,,\\
\mathrm{U(3)_L} &:\qquad 3n_{Q_{2,3}}-3n_{Q_1}-n_{L}=6-3-3=0\,,\\
\mathrm{U}(1)   &:\qquad 3n_{L}-n_{R}-n_{S}=9-3-6=0\,,\\
\mathrm{U}(1)'  &:\qquad 3n_{D}-3n_{U}+n_{R}-n_{S}=15-12+3-6=0\,.
\end{split}
\ee

The second important condition observed by the quiver field theory reads
\be{condition2}
\sum_{a}C_{a}N_{a}(\pi_{a}-\pi'_{a})=0\,, \ee
and guarantees the existence of a massless vector boson associated
with the unbroken $\mathrm{U(1)_{X}}=\sum_{a}C_{a}\mathrm{U}(1)_{a}$
symmetry \cite{Aldazabal:2000dg,Cvetic:2009yh,Cvetic:2011iq}.  Again,
in terms of field representations, Eq.~(\ref{condition2}) can be
written as
 \be{condition3a}
C_{a}N_{a}(\#S_{a}-\# \bar{S}_{a}+\# A_{a}-\# \bar{A}_{a})-\sum_{b\neq a} C_{b}N_{b}\left[\#(F_a,\bar{F}_b) - \#(F_a,F_b)\right]=0\,,
 \ee
and is satisfied by $\mathrm{U(1)_{X}}$ accordingly:

\be{tpc2}
\begin{split}
\mathrm{3_c} &:\qquad -3 C_{\rm 3_L}\left(n_{Q_1}-n_{Q_{2,3}}\right)-C_{\rm 1'}\left(-n_{D}+n_{U}\right)=\frac{1}{2}\left(1-2\right)-\frac{1}{2}\left(-5+4\right)=0\,,\\
\mathrm{3_L} &:\qquad -3 C_{\rm 3_c}\left(-n_{Q_1}-n_{Q_{2,3}}\right)-C_{\rm 1}\left(-n_{L}\right)=-\frac{1}{2}\left(-1-2\right)+\frac{1}{2}\left(-3\right)=0\,,\\
\mathrm{1} &:\qquad -3 C_{\rm 3_L}\left(n_{L}\right)-C_{\rm 1'}\left(-n_{R}+n_{S}\right)=\frac{1}{2}(3)-\frac{1}{2}\left(-3+6\right)=0\,,\\
\mathrm{1'} &:\qquad -3 C_{\rm 3_c}\left(n_{D}+n_{U}\right)-C_{\rm 1}\left(n_{R}+n_{S}\right)=-\frac{1}{2}(5+4)+\frac{1}{2}\left(3+6\right)=0\,.\\
\end{split}
\ee

We conclude this section pointing out that the remaining Abelian and
mixed anomalies can be canceled by a Generalized Green-Schwarz
mechanism with Stueckelberg, axionic and generalized Chern-Simons
couplings, following the lines of
\cite{Coriano':2005js,Anastasopoulos:2006cz,Anastasopoulos:2008jt}. This
mechanism introduces non trivial interactions among the various gauge
bosons of the model and provides potentially interesting
phenomenological implications.

\section{Phenomenology}

\subsection{Quark sector}

Fermion masses are obtained perturbatively, after spontaneous breaking
of the gauge symmetry, from the Yukawa interactions present in
Eq.~(\ref{super}). For the quarks one has the following mass matrices

\begin{equation}
m_u=\frac{1}{\sqrt{2}}\left(
\begin{array}{cccc}
 k'_2 y_5^{uu}+k'_3 y_6^{uu} & k'_2 y_5^{uc}+k'_3 y_6^{uc} & k'_2 y_5^{ut}+k'_3 y_6^{ut} & k'_2 y_5^{uu'}+k'_3
   y_6^{uu'} \\
 k_1 y_7^{cu} & k_1  y_7^{cc} & k_1  y_7^{ct} & k_1  y_7^{cu'} \\
 k_1 y_7^{tu} & k_1  y_7^{tc} & k_1  y_7^{tt} & k_1  y_7^{tu'} \\
 n'_2 y_5^{uu}+n'_3 y_6^{uu} & n'_2 y_5^{uc}+n'_3 y_5^{uc} & n'_2 y_5^{ut}+n'_3 y_5^{ut} & n'_2 y_5^{uu'}+n'_3
   y_5^{uu'} \\
\end{array}
\right)\,,
\end{equation}

\begin{equation}
m_d=\frac{1}{\sqrt{2}}\left(
\begin{array}{ccccc}
 k'_1 y_4^{dd} & k'_1 y_4^{ds} & k'_1 y_4^{db} & k'_1 y_4^{ds'} & k'_1 y_4^{db'} \\
 k_2 y_8^{sd}+ k_3 y_9^{sd} & k_2 y_8^{ss}+ k_3 y_9^{ss} & k_2 y_8^{sb}+ k_3 y_9^{sb} & k_2 y_8^{ss'}+k_3  y_9^{ss'} & k_2 y_8^{sb'}+ k_3 y_9^{sb'}\\
 k_2 y_8^{bd}+ k_3 y_9^{bd} & k_2 y_8^{bs}+ k_3 y_9^{bs}& k_2 y_8^{bb}+ k_3 y_9^{bb} & k_2 y_8^{bs'}+ k_3 y_9^{bs'} & k_2 y_8^{bb'}+ k_3 y_9^{bb'} \\
 n_2 y_8^{sd}+ n_3 y_9^{sd} & n_2 y_8^{ss}+ n_3 y_9^{ss} & n_2 y_8^{sb}+ n_3  y_9^{sb} & n_2 y_8^{ss'}+ n_3 y_9^{ss'} & n_2 y_8^{sb'}+ n_3 y_9^{sb'} \\
 n_2 y_8^{bd}+ n_3 y_9^{bd} & n_2 y_8^{bs}+ n_3 y_9^{bs} & n_2 y_8^{bb}+ n_3 y_9^{bb} & n_2 y_8^{bs'}+ n_3 y_9^{bs'} & n_2 y_8^{bb'}+ n_3 y_9^{bb'} \\
\end{array}
\right)\,,
\end{equation}
where we have assumed that the scalar fields
$\phi^{(\prime)}_{1,\,2,\,3}$ contained in $\Phi^{(\prime)}_{1,\,2,\,3}$, develop vacuum expectation values in all neutral directions:
\begin{equation}
\langle\phi^{(\prime)}_{1}\rangle=\frac{1}{\sqrt{2}}\left(\begin{array}{c}
k^{(\prime)}_1\\0\\0
\end{array}\right)\,,\qquad
\langle\phi^{(\prime)}_{2,3}\rangle=\frac{1}{\sqrt{2}}\left(\begin{array}{c}
0\\k^{(\prime)}_{2,3}\\n^{(\prime)}_{2,3}
\end{array}\right)\,.
\end{equation}
Here $n^{(\prime)}_{1,2,3}$ characterizes the $\mathrm{SU(3)_L}$ breaking and $k^{(\prime)}_{1,2,3}$ the subsequent $\mathrm{SU(2)_L}$ breakdown.

One can verify that a realistic pattern of quark masses and
interactions can be obtained from the above mass matrices, though its
detailed study is beyond the scope of the present paper.
One characteristic feature which we can comment is the existence of
heavy exotic quarks which, in general, mix with those of the \SM
leading to an effective violation of unitarity of the CKM
matrix~\cite{Botella:2008qm,Morisi:2013eca}.
Furthermore, the presence of heavy exotic quarks may lead to a
number of phenomenological implications, such as accommodating the
recent diphoton anomaly~\cite{Boucenna:2015pav,Dong:2015dxw}.

\subsection{Neutrino  masses}

After spontaneous symmetry breaking, one obtains a Dirac neutrino
mass~\cite{Valle:2016kyz}
\begin{equation}\label{nu_mass}
\begin{split}
-\mathcal{L}_{\rm mass}=\frac{1}{\sqrt{2}}\left(\begin{array}{cc}
\bar{\nu}_L & \bar{N}_L
\end{array}  \right) 
\left(\begin{array}{cc}
y_{2} k'_2+y_{3} k'_3 & \tilde{y}_{2} k'_2+\tilde{y}_{3} k_3\\
y_{2} n'_2+y_{3} n'_3 & \tilde{y}_{2} n'_2+\tilde{y}_{3} n_3\\
\end{array}\right)
\left(\begin{array}{c}
S_R\\
\tilde{S}_R
\end{array}  \right) + \mathrm{h.c.},
\end{split}
\end{equation}
where $y_{2,3}$ and $\tilde{y}_{2,3}$ are $3\times 3$ Yukawa matrices and we have denoted $S=S_{1\cdots 3}$ and $\tilde{S}=S_{4\cdots 6}$.
The light neutrino masses can be readily estimated in the one family
approximation which, as usual, is diagonalized by a bi-unitary
transformation $\mathcal{M}_{\rm
  diag}=U_{\nu}^{\dagger}\mathcal{M}U_{S}$.
As can be seen, the effective light neutrino mass vanishes as the scalar
vacuum expectation values $n_{2,3}$ become large with respect to $|k'_2 n'_3 - k'_3 n'_2|$, very much as expected in
the conventional Majorana neutrino seesaw mechanism.

Another feature is that the light neutrino become massless in the
limit where the dynamical alignment parameter~\cite{Valle:2016kyz}
$$k'_2 n'_3 - k'_3 n'_2$$
approaches zero.
The same holds for the up quark.  Hence, in the present formulation,
the same alignment yields a massless $u$ quark, implying a tension
between small neutrino masses and a realistic $u$ quark. However, this tension is
still comparable with {\it e.g.} the Yukawa hierarchy between the
electron and the top quark in the \sm  model.
Moreover, for typical parameter choices, this requires an upper limit
to the new scale associated to the new gauge bosons, and the need for
an adequate choice of Yukawa coupling parameters in order to attain a
realistic mass pattern in both quark and lepton sectors.

We conclude this section pointing out that a third interesting feature of Eq. (\ref{nu_mass}) is the fact that the light neutrino mass also vanish in the limit of democratic Yukawa couplings $y_2=\tilde{y}_2$, $y_3=\tilde{y}_3$. A discrete symmetry favoring this relation can be implemented in a more complete model in order to relieve the tension between the neutrino and up quark scales.

\subsection{Exotic instantons and $n-\bar{n}$ oscillations}

\begin{figure}[t]
\centerline{ \includegraphics [width=0.9 \columnwidth]{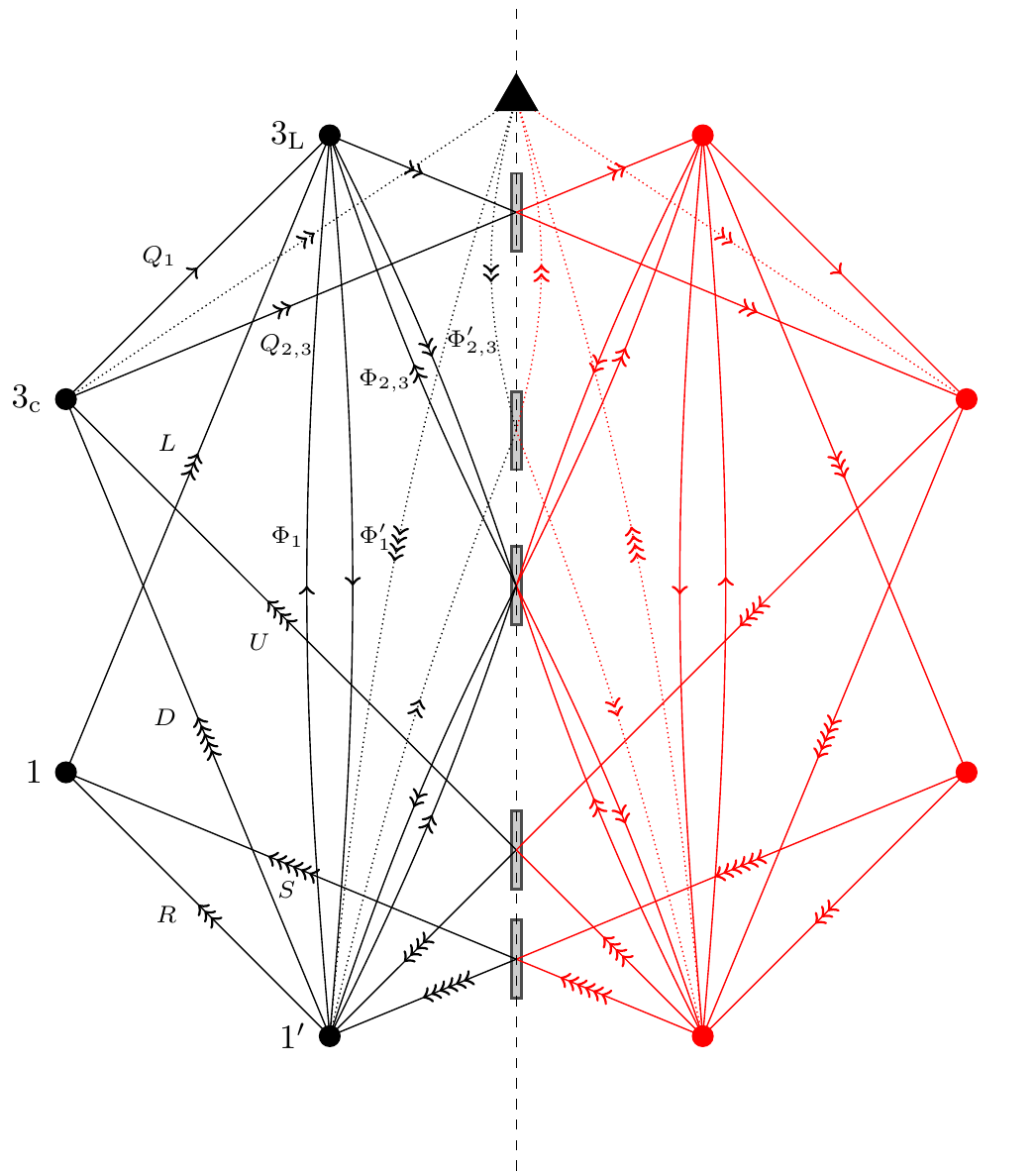}}
\vspace*{-1ex}
\caption{ \u3u311 modified (un)oriented quiver theory with an intersecting E2-brane.}
\label{plot2}   
\end{figure}

In this section we discuss the possibility of inducing
neutron-antineutron oscillations by exotic instanton effects generated
by the action of E2-branes intersecting other physical D6-branes in
our \u3u311 model.  Fig.~\ref{plot2} shows the modified setup obtained
by adding an E2-brane with the following set of intersections: two
with the $\mathrm{U(3)_{c}}$-stack, two with the
$\mathrm{U(1)'}$-stack, and four with the $\mathrm{\hat{U}(1)'}$-stack
(image of $\mathrm{U(1)'}$). There are 3 different species of modulini
associated with each intersection of the E2-brane and the D6-branes of
the model. We denote by $\tau^{i}$ those living between
E2$-\mathrm{U(3)_{c}}$, $\alpha$ between E2$-\mathrm{\hat{U}(1)'}$,
and $\beta$ between E2$-\mathrm{U(1)'}$.  The effective interactions
among the quarks $U$, $D$ and the modulini are given by:
\be{effectivemod} \mathcal{L}_{eff}\sim
K_{f}^{(1)}U_{f}^{i}\tau_{i}\alpha+K_{f'}^{(2)}D_{f'}^{i}\tau_{i}\beta
\,, \ee where $f$ and $f'$ stand for the flavor indices of the
corresponding fields and $i$ is the $\mathrm{U(3)_{c}}$ index.
Integrating over the modulini space associated to the D6-E2
intersections, we obtain \be{obtain} \mathcal{W}_{D6-\hat{D6}-E2}=\int
d^{6}\tau d^{4}\beta d^{2}\alpha e^{\mathcal{L}_{eff}}=
\mathcal{Y}\frac{e^{-S_{E2}}}{M_{S}^{3}}\epsilon_{ijk}\epsilon_{i'j'k'}U^{i}D^{j}D^{k}U^{i'}D^{j'}D^{k'}\,,
\ee with the flavor matrix $\mathcal{Y}_{f_{1}f_{2}f_{3}f_{4}f_{5}f_{6}}\equiv
K^{(1)}_{f_{1}}K^{(1)}_{f_{2}}K^{(2)}_{f_{3}}K^{(2)}_{f_{4}}K^{(2)}_{f_{5}}K^{(2)}_{f_{6}}$. As the coefficients $K^{(1,2)}$ parametrize
particular homologies of the mixed disk amplitudes, we treat them as
free parameters, since our model is local. Thus, the superpotential
(\ref{obtain}) leads to an effective dimension 9 six-quark operator
$\mathcal{O}_{n\bar{n}}=(u^{c}d^{c}d^{c})^{2}/\mathcal{M}^{5}$
responsible for neutron-antineutron oscillations. The related new
physics scale $\mathcal{M}$ can be written as
$\mathcal{M}^{5}=y_{1}^{-1}e^{+S_{E2}}M_{S}^{3}m_{\tilde{g}}^{2}$,
where $m_{\tilde{g}}$ is determined by gaugino-mediated quark-squark
SUSY reductions (see \cite{Addazi:2015ewa} for example), and
$y_{1}\equiv\mathcal{Y}_{111111}$.

In terms of $\mathcal{M}$, the $n-\bar{n}$ transition time (in vacuum)
reads $\tau_{n\bar{n}}\simeq \mathcal{M}^{5}/\Lambda_{QCD}^{6}$.  The
current bounds on the $n-\bar{n}$ transition time are
$\tau_{n\bar{n}}\gtrsim 3\, \rm yrs$, constraining the new physics
scale to $\mathcal{M}\gtrsim 300\, \rm TeV$.  The next generation of
experiments is expected to test $\mathcal{M}\simeq 1000\, \rm TeV$
\cite{Phillips:2014fgb}, an interesting scale that can be reproduced
by different choices of parameters
$(y_{1},e^{+S_{E2}},M_{S},m_{\tilde{g}})$.  For example, one can
envisage a scenario in which $y_{1}^{-1}e^{+S_{E2}}\simeq 1$,
$m_{\tilde{g}}\simeq 1\, \rm TeV$, $M_{S}\simeq 10^{5}\, \rm TeV$,
compatible with TeV-scale supersymmetry and related to the naturalness
of the Higgs mass.  Alternatively, the same scale can be achieved by
an unnatural scenario with $m_{\tilde{g}}\simeq M_{S}\simeq 1000\, \rm
TeV$ ($y_{1}^{-1}\simeq e^{+S_{E2}}\simeq 1$). In the latter case, the
hierarchy problem is not solved but it is strongly alleviated by
virtue of a low string scale, {\it i.e.}  the original hierarchy of
$m_{H}^{2}/M_{Pl}^{2}\simeq 10^{-34}$ is reduced to merely $10^{-8}$.

\section{Conclusions}

In this paper, we have proposed a consistent ultra-violet completion
of a \SM extension based on the gauge symmetry \s3s31 within the
context of an open string theory. 
In particular, we have constructed an example of such a $331$ model in
which the following properties are satisfied in a consistent way: i)
all stringy tadpoles as well as anomalous $\mathrm{U(1)}$s are
avoided; ii) all desired Yukawa terms are allowed at the perturbative
level by open string orientations, which are in turn generated by
non-perturbative effects; iii) R-parity is preserved automatically at
tree-level, avoiding proton destabilization and other undesired
operators; iv) Dangerous contributions to couplings of
$\mathrm{U(1)_{X}}$-RR fields cancel consistently by virtue of
Eq.(\ref{tpc}). 

Even though our 331 gauge model lacks an embedding
  into a conventional unified field theory, we show here how it is
  nicely embedded in a quiver theory, free of gauge and stringy
  anomalies. In such construction lepton and baryon numbers are
  conserved at the perturbative level, so neutrinos are Dirac
  particles.  Dynamical violation of baryon/lepton numbers can be
  introduced through exotic instanton effects. We have studied a
  particular setup for the generation of non-perturbative $\Delta B=2$
  violating operators which would lead to neutron-antineutron
  oscillations and possible collider signatures.  In contrast, proton
  decay and other dangerous R--parity violating processes are
  forbidden.

Finally, our quiver theory suggests the presence extra observables
peculiar of string theories. For example in the context of a low scale
string theory $M_{S}=10\div 10^{5}\, \rm TeV$, one may have extra
anomalous heavy neutral Abelian bosons, interacting through
generalized Chern-Simons terms with the $331$ neutral gauge bosons
$\gamma,\,Z,\,Z',\,X$. In addition there is the exciting possibility
of finding direct signatures of higher-spin resonances in future
colliders beyond LHC.

\vspace{1cm} 

{\large \bf Acknowledgments} 
\vspace{3mm}

Work supported by the Spanish grants FPA2014-58183-P, Multidark
CSD2009-00064 and SEV-2014-0398 (MINECO), and PROMETEOII/2014/084
(Generalitat Valenciana).  Work of A.~A was supported in part by the
MIUR research grant ``Theoretical Astroparticle Physics'' PRIN
2012CPPYP7.  C.A.V-A. acknowledges support from CONACYT (MEXICO),
grant 251357.



\end{document}